\documentclass[preprint2]{emulateapj}
\usepackage{natbib}
\usepackage{hyperref}
\usepackage{breakurl}
\bibpunct[; ]{(}{)}{,}{a}{}{;}

\newcommand{\ha}{H$\alpha$}
\newcommand{\lya}{Ly$\alpha$}
\newcommand{\lyb}{Ly$\beta$}
\newcommand{\lyg}{Ly$\gamma$}
\newcommand{\lyd}{Ly$\delta$}
\newcommand{\lye}{Ly$\epsilon$}

\shorttitle{Detection of 3-Minute Oscillations in Full-Disk \lya\ Emission During A Solar Flare}

\shortauthors{Milligan et al.}

\begin{document}

\title{Detection of 3-Minute Oscillations in Full-Disk \lya\ Emission During A Solar Flare}

\notetoeditor{the contact email is ryan.milligan@glasgow.ac.uk and is the only one which should appear on the journal version}

\author{Ryan O. Milligan\altaffilmark{1,2,3,4}, Bernhard Fleck\altaffilmark{5}, Jack Ireland\altaffilmark{3,6}, Lyndsay Fletcher\altaffilmark{1}, \& Brian R. Dennis\altaffilmark{3}}

\altaffiltext{1}{School of Physics and Astronomy, University of Glasgow, Glasgow, G12 8QQ, UK}
\altaffiltext{2}{Astrophysics Research Centre, School of Mathematics \& Physics, Queen's University Belfast, University Road, Belfast, Northern Ireland, BT7 1NN}
\altaffiltext{3}{Solar Physics Laboratory (Code 671), Heliophysics Science Division, NASA Goddard Space Flight Center, Greenbelt, MD 20771, USA}
\altaffiltext{4}{Department of Physics Catholic University of America, 620 Michigan Avenue, Northeast, Washington, DC 20064, USA}
\altaffiltext{5}{ESA Directorate of Science, Operations Department, c/o NASA/GSFC Code 671, Greenbelt, MD 20071, USA}
\altaffiltext{6}{ADNET Systems, Inc.}

\begin{abstract}
In this Letter we report the detection of chromospheric 3-minute oscillations in disk-integrated EUV irradiance observations during a solar flare. A wavelet analysis of detrended Lyman-alpha (from GOES/EUVS) and Lyman continuum (from SDO/EVE) emission from the 2011 February 15 X-class flare (SOL2011-02-15T01:56) revealed a $\sim$3-minute period present during the flare's main phase. The formation temperature of this emission locates this radiation to the flare's chromospheric footpoints, and similar behaviour is found in the SDO/AIA 1600\AA\ and 1700\AA\ channels, which are dominated by chromospheric continuum. The implication is that the chromosphere responds dynamically at its acoustic cutoff frequency to an impulsive injection of energy. Since the 3-minute period was not found at hard X-ray energies (50--100~keV) in RHESSI data we can state that this 3-minute oscillation does not depend on the rate of energization of non-thermal electrons. However, a second period of 120~s found in both hard X-ray and chromospheric emission is consistent with episodic electron energization on 2-minute timescales. Our finding on the 3-minute oscillation suggests that chromospheric mechanical energy should be included in the flare energy budget, and the fluctuations in the Lyman-alpha line may influence the composition and dynamics of planetary atmospheres during periods of high activity.
\end{abstract}


\section{INTRODUCTION}
\label{intro}
Quasi-periodic pulsations (QPPs) are widely reported in emission from solar flares. These are regular fluctuations in the flare radiation intensity, which are very clear in hard X-rays (HXRs) and radio waves generated by non-thermal electrons (\citealt{naka09}, \citealt{vand16} and \citealt{ingl16}), but also detected over a wide range of wavelengths\footnote[2]{See \url{http://www2.warwick.ac.uk/fac/sci/physics/research/cfsa/people/valery/research/qpp/} for a comprehensive list of relevant publications, and see \url{https://aringlis.github.io/AFINO/} for an automatically updated list of GOES SXR QPP observations.}. QPPs in non-thermal signatures are widely interpreted as either revealing the magneto-hydrodynamic (MHD) oscillation modes of the flare's magnetic environment, or reflecting an oscillatory driver for electron acceleration. There have been relatively few reports of quasi-periodic behavior in the emission from the thermal plasmas of the solar atmosphere, including the chromosphere, in response to flare energization. QPPs with periods of $\lesssim$1~minute have been found in UV/EUV/SXR flare emission (e.g. \citealt{doll12}, \citealt{simo15}) perhaps signaling MHD oscillations in post-flare coronal loops, or coronal loop-filling by heated plasma expanding from a periodically-heated chromosphere. \cite{bros16} found a period of 171~s in Interface Region Imaging Spectograph (IRIS; \citealt{depo14}) observations of chromospheric flare \ion{C}{1} line emission, and interpreted it as chromospheric heating due to quasi-periodic injection of non-thermal electrons. Other studies have found fluctuations of 1--4~minutes in flare chromospheric emission that are interpreted as evidence for episodic reconnection driven by leakage of slow-mode oscillations from an underlying sunspot \citep{sych09,li15,kuma16,ning17}. 

Different pulsation periods may be present at different phases of the flare. For example, \cite{haye16} found that in the X-class flare SOL2013-10-28 the period of the observed radio and X-ray QPPs changes with short period pulsations ($\sim$40s) dominating during the impulsive, energy-release phase, and longer-period pulsations ($\sim$80s) more prevalent during the gradual, decay phase. Similarly \cite{denn17} found the period of QPPs to change from $\sim$25~s to $\sim$100~s during an the X-class flare SOL2013-05-14. This may reflect a change in the dominant driver of pulsations, e.g. from periodic electron acceleration during the impulsive phase, to MHD oscillations of hot post-flare loops in the decay phase.

It is well known that the (non-flaring) solar chromosphere oscillates with a dominant period of around 3 minutes. The decrease of the average period of oscillations in the solar atmosphere from about 5 minute in the photosphere to 3 minute in the chromosphere (e.g. \citealt{noye63}) is due to the strong spatial damping of evanescent waves with height, whereas the 3 minute oscillations at the cut-off frequency are not damped. \cite{flec91}, \cite{kalk94}, \cite{sutm98}, and \cite{chae15} and others have shown theoretically that any disturbance in the chromosphere, whether impulsive or quasi-periodic, causes it to oscillate at its acoustic cutoff frequency  (the ``Lamb effect'';  \citealt{lamb09}).  For the chromosphere this cutoff frequency is 5.5~mHz, corresponding to a 3-minute period. Convincing observational evidence for impulsive excitation of oscillations at the acoustic cutoff frequency was recently presented by \cite{kwak16} who detected periods of 2.7--3.3~minutes in response to a strong downflow event detected in \ha+0.5\AA. This was seen in spatially-resolved chromospheric (\ion{Mg}{2}, \ion{Ca}{2}) and transition region (\ion{C}{2}, \ion{Si}{4}) lines measured by IRIS. The authors concluded that these oscillations represent gravity-modified acoustic waves generated by an impulsive disturbance in the chromosphere. 

\begin{figure}[!t]
\begin{center}
\includegraphics[width=0.5\textwidth]{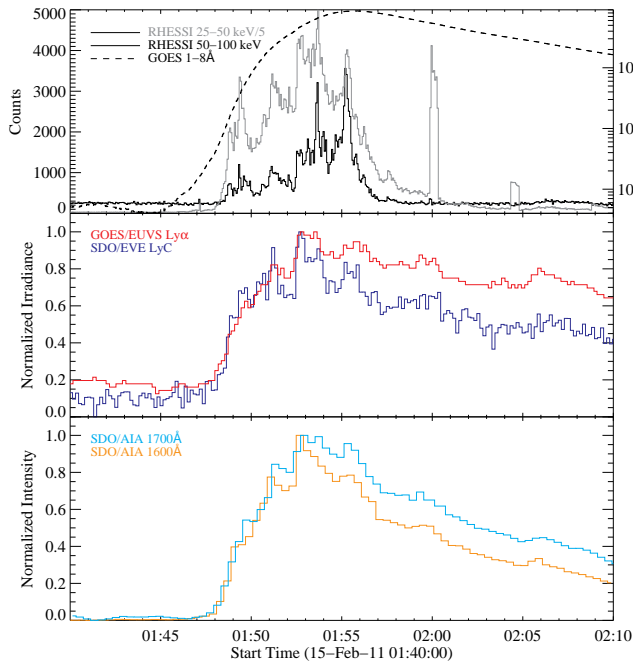}
\caption{Plot of lightcurves of different emission during the 2011 February 15 flare. Top panel: RHESSI 50-100~keV (black), 25--50~keV (grey; scaled by a factor of 5), and GOES 1--8\AA\ (dashed curve). Spikes at 02:00~UT and 02:05~UT in 25--50~keV are due to attenuator state changes. Middle panel: Lyman-alpha from GOES/EUVS (red) and Lyman continuum from SDO/EVE (blue). Bottom panel: UV continua from SDO/AIA (1600\AA\ - orange; 1700\AA\ - cyan).}
\label{fig:goes_hsi_lya_lyc_aia_ltc}
\end{center}
\end{figure}

In this Letter we present evidence that oscillations with periods around 3 minutes are present in the chromospheric emission during the impulsive phase of an X-class solar flare. We show that the 3-minute period is very prominent in the hydrogen Lyman-alpha line, and the hydrogen Lyman continuum, both characteristic of chromospheric plasma at around 10,000K. However, the 3-minute period is not seen in HXRs (which show a dominant 2-minute signal) indicating that this 3-minute signature is not due to quasi-periodic electron injection. In Section~\ref{sec:data_anal}, the datasets used and the analysis techniques employed are described. Section~\ref{sec:results} outlines the findings, while Section~\ref{sec:conc} provides some discussion and interpretation

\section{Observations and Data Analysis}
\label{sec:data_anal}

One of the most studied solar flares of Solar Cycle 24 is the X2.2 flare that occurred on 2011 February 15 (SOL2011-02-15T01:56). It was the first X-class flare of the cycle and was observed by a number of different instruments. The top panel of Figure~\ref{fig:goes_hsi_lya_lyc_aia_ltc} shows the lightcurves of 25--50 and 50--100~keV hard X-ray (HXR) emission from the Reuven Ramaty High Energy Solar Spectroscopic Imager (RHESSI; \citealt{lin02}) at 4~second cadence, and of 1--8~\AA\ SXR emission from the X-Ray Sensor (XRS; \citealt{hans96}; dashed curve) on the Geostationary Operational Environmental Satellite (GOES15; \citealt{vier07}) at 2~second cadence, for 30~minutes around the rise and peak of the X-class flare. The middle panel shows the chromospheric response in both the Lyman-alpha line at 1216\AA\ (hereafter, \lya; red curve) from the E-channel of the EUV Sensor (EUVS-E) on GOES, and the Lyman continuum blue-ward of 912\AA\ (hereafter, LyC; blue curve) from the EUV Variability Experiment (EVE; \citealt{wood12}) on the Solar Dynamics Observatory (SDO; \citealt{pesn12}) at 10.24~s and 10~s cadence, respectively. For comparison, the time series of 1600\AA\ and 1700\AA\ emission from the Atmospheric Imaging Assembly (AIA; \citealt{leme12}), also on SDO, are  shown in the bottom panel. These data were taken at 24~s cadence. During this event both channels saturated around the peak of the flare. Following \cite{mill14}, a 200''$\times$200'' area around the flare site was integrated over in each channel to ensure no loss of counts when deriving the lightcurves.

\begin{figure}[!t]
\begin{center}
\includegraphics[width=0.5\textwidth]{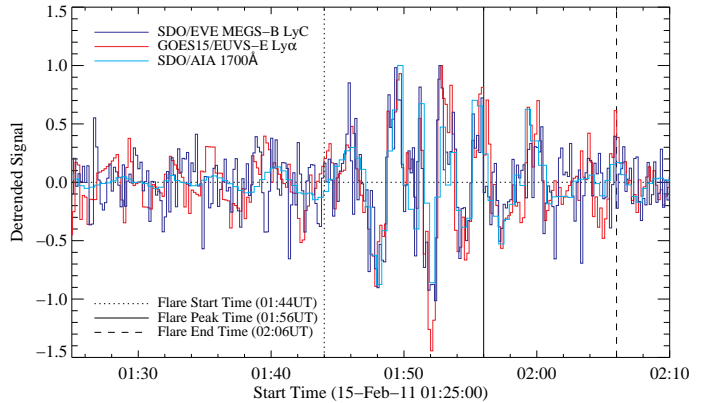}
\caption{Plot of the periodic behavior of chromospheric emission: \lya\ (red), LyC (blue), and 1700\AA\ (cyan). The data have been detrended using an FFT filter with a cutoff period of 400~s in each case. The vertical dotted, solid, and dashed lines denote the start, peak, and end times of the flare, respectively, as determined by the 1--8\AA\ channel of GOES/XRS.}
\label{fig:goes_eve_aia_qpp}
\end{center}
\end{figure}

\begin{figure*}[!t]
\begin{center}
\includegraphics[width=0.49\textwidth]{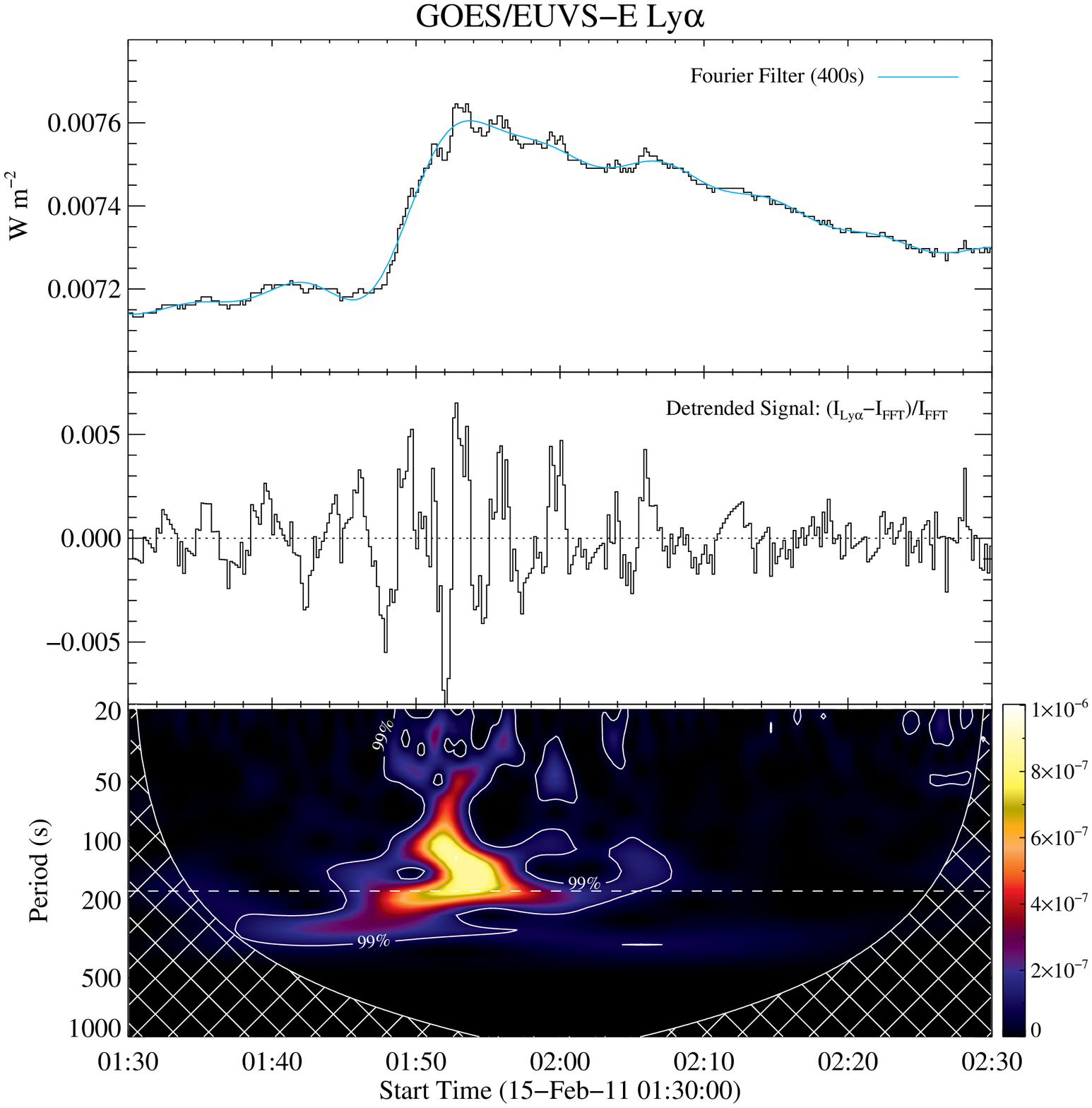}
\includegraphics[width=0.49\textwidth]{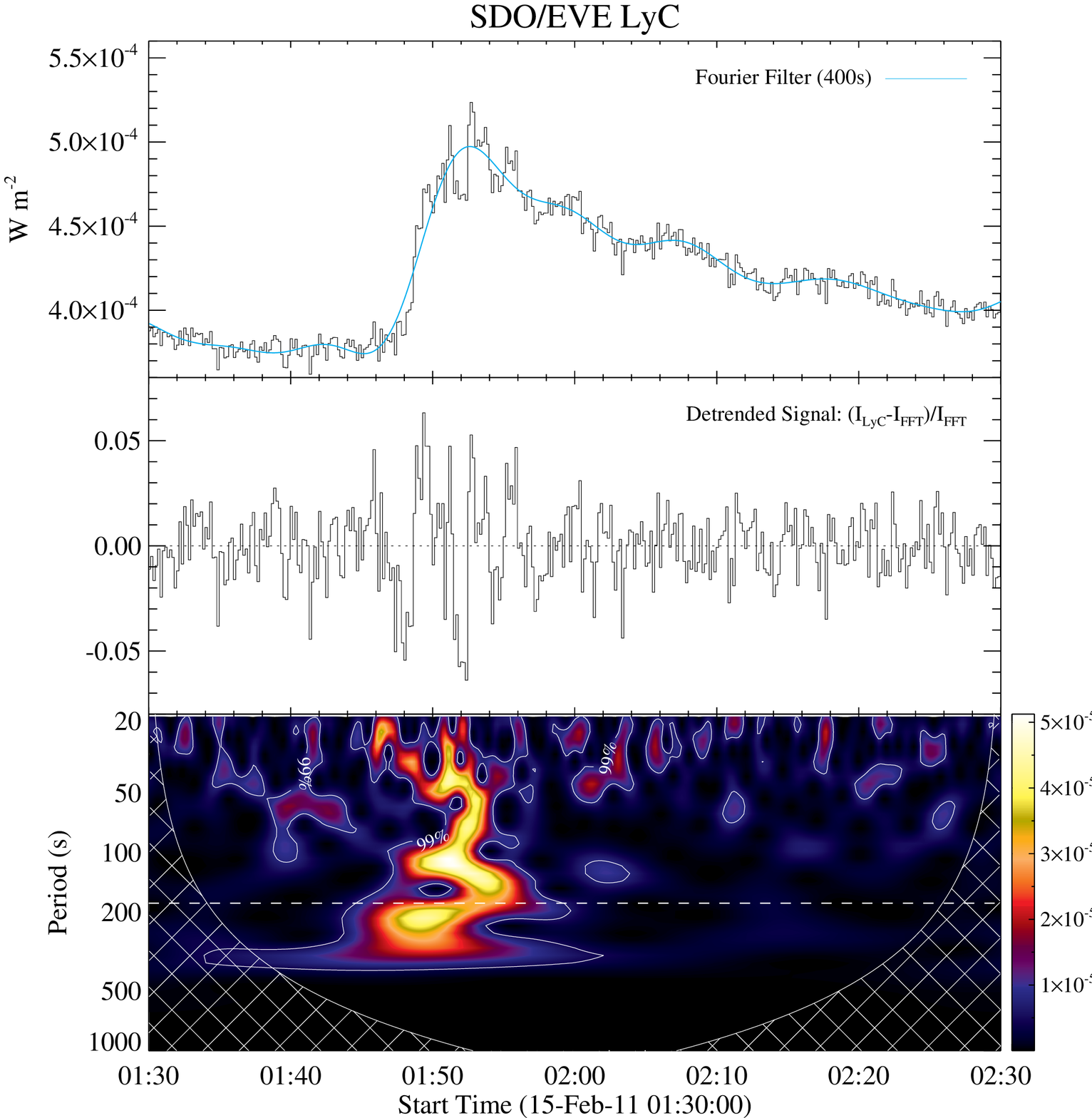}
\caption{Analysis of the \lya\ (left) and LyC (right) emission during the 2011 February 15 flare. Top panels: full-disk irradiance lightcurves with a 400s FFT high-pass filter overlaid in cyan. Middle panels: the detrended time profiles after subtracting the filtered profiles. Bottom panels: Wavelet spectra (power per unit time) of the detrended profiles with 99\% significance levels overlaid. The horizontal dashed white line denotes a period of 180~s.}
\label{fig:lya_lyc_qpp}
\end{center}
\end{figure*}

The SDO/EVE MEGS-P photometer is intended to record \lya\ irradiance, but \cite{mill16} have identified unexpected behavior in this signal. The EUV Sensor (EUVS) on GOES15 was therefore used instead. The E-channel on EUVS spans the \lya\ line in a broadband ($\sim$100\AA) manner similar to MEGS-P, and was operational during the 2011 February 15 flare. The \lya\ lightcurves plotted in the middle panel of Figure~\ref{fig:goes_hsi_lya_lyc_aia_ltc} were generated using Version 4 of the data. This version of the data has been corrected for degradation and calibrated using SORCE/SOLSTICE \lya\ measurements.

The MEGS-B (Multiple EUV Grating Spectrograph) component of EVE obtains spectra over the 370--1050\AA\ range at 10~s cadence and 1\AA\ resolution. Aside from the numerous spectral lines that occupy this wavelength range, the most prominent feature is that of the free--bound LyC with a recombination edge at 912\AA. Time profiles of this continuum emission during flares from EVE were first presented by \cite{mill12}, and followed up by a study of the energetics of this and other chromospheric emissions during the 2011 February 15 flare \citep{mill14}. In order to isolate the continuum emission from the overlying emission lines, \cite{mill14} employed a RANdom Sample Consensus (RANSAC; \citealt{fisc81}) technique that treats the lines as outliers over a chosen wavelength range (see appendix of \citealt{mill14} for more details). Integrating under this fit at each time step allowed the lightcurve of LyC in middle panel of Figure~\ref{fig:goes_hsi_lya_lyc_aia_ltc} to be derived. The same technique was applied to Version 5 of the EVE data in this study.

To highlight the periodic behavior in the \lya\ and LyC emissions, the lightcurves were detrended using a Fast Fourier Transform (FFT) filter. A cutoff period of 400~s (2.5~mHz) was chosen for this analysis, but the choice of frequency was not found to affect the derived period (see 
Section~\ref{sec:results}). The resulting periodic behavior in both \lya\ and LyC emission is shown in Figure~\ref{fig:goes_eve_aia_qpp} (red and blue curves, respectively). The detrended 1700\AA\ profile is also shown for comparison. There is a remarkable agreement between all three detrended profiles in terms of the phase of each pulsation. The coincidence between profiles taken by three different instruments implies that the bursts are genuinely solar in origin, and that they originate in the chromosphere. The similarity is further evidence that the spikes seen in LyC are not an artifact of the fitting algorithm. 

The raw lightcurves of \lya\ and LyC emission are shown in the top panels of Figure~\ref{fig:lya_lyc_qpp}. Overplotted are the low-pass (400~s cutoff period) Fourier filtered time series in cyan. Having removed the smoothly varying component of the flare time profile for both \lya\ and LyC, the resulting detrended profiles are shown in the middle panels of Figure~\ref{fig:lya_lyc_qpp}. The final step was to apply the standard wavelet analysis technique of \cite{torr98} to determine the period(s) of the pulsations during the flare. The corresponding wavelet spectra per unit time are shown in the bottom panels of Figure~\ref{fig:lya_lyc_qpp}.

\section{Results}
\label{sec:results}

\begin{figure*}[!t]
\begin{center}
\includegraphics[width=0.65\textwidth]{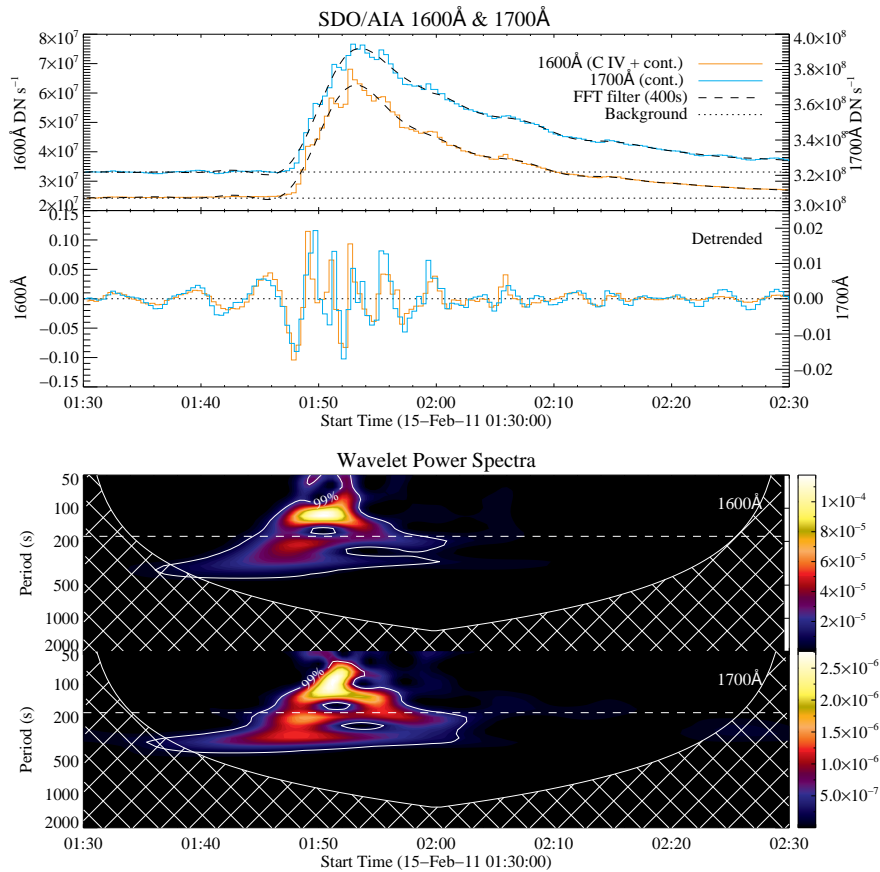}
\caption{Top panels: lightcurves of 1600\AA\ (orange) and 1700\AA\ (cyan) emission obtained by integrating over a 200$\arcsec$ $\times$ 200$\arcsec$ field of view centered on NOAA AR 11158 in SDO/AIA images (upper) and the corresponding detrended profiles with a 400~s cutoff period (lower). Bottom panels: Wavelet power spectra for 1600\AA\ (upper) and 1700\AA\ (lower) with 99\% significance levels overlaid. Horizontal dashed lines denote a period of 180~s.}
\label{fig:aia_uv}
\end{center}
\end{figure*}

The wavelet power spectrum for the detrended \lya\ time series in the bottom left panel of Figure~\ref{fig:lya_lyc_qpp} shows enhanced power over a broad range of periods during the rise and peak of the flare (01:45--02:00~UT). The bulk of this power is evident at periods around 100-200~s. There is also a similar distribution of power in frequency and time in the LyC spectrum (bottom right panel of Figure~\ref{fig:lya_lyc_qpp}). Both spectra also show enhanced power at higher frequencies around the time of the flare onset. The enhanced power around 180~s (5.5~mHz; horizontal dashed white lines in both bottom panels) corresponds to the acoustic cutoff frequency in the chromosphere. The 180~s period is not apparent in the quiescent, full-disk signal from GOES/EUVS-E or SDO/EVE, presumably due to the incoherence of the signal in disk-integrated emission, although longer period (300~s) oscillations are apparent in non-flaring regions of AIA 1600\AA\ and 1700\AA\ images (see below). The flare therefore seems to either initiate the oscillation itself, or it amplifies or enhances a pre-existing oscillation. In the latter case, the flare may either drive a pre-existing dynamical behavior, or it changes the properties of the radiating gas so that the variations in intensity become more visible (see Section~\ref{sec:conc} for further discussion). It is also worth noting that this 3-minute power is readily apparent in wavelet analysis of the raw \lya\ data with no detrending applied.

The two UV channels on SDO/AIA - 1600\AA\ and 1700\AA\ - image the solar chromosphere at 24~s cadence. While this emission is largely continuum \citep{leme12} rather than hydrogen line emission, it is worth including given that any flare emission (ribbons) should come from the same spatial location as the \lya\ and LyC emission, although they may originate at different depths in the flaring atmosphere. These data were again detrended using an FFT filter (top panels of Figure~\ref{fig:aia_uv}). The bottom two panels of Figure~\ref{fig:aia_uv} show the resulting wavelet power spectra for the two channels. The 3-minute oscillation is apparent, as well as even stronger power at $\sim$120~s. Evidence for 3-minute oscillations was also found in the \lyb\ line from MEGS-B, although no such oscillations were detected in the higher order Lyman lines (\lyg, \lyd, \lye).

\begin{figure*}[!t]
\begin{center}
\includegraphics[width=\textwidth]{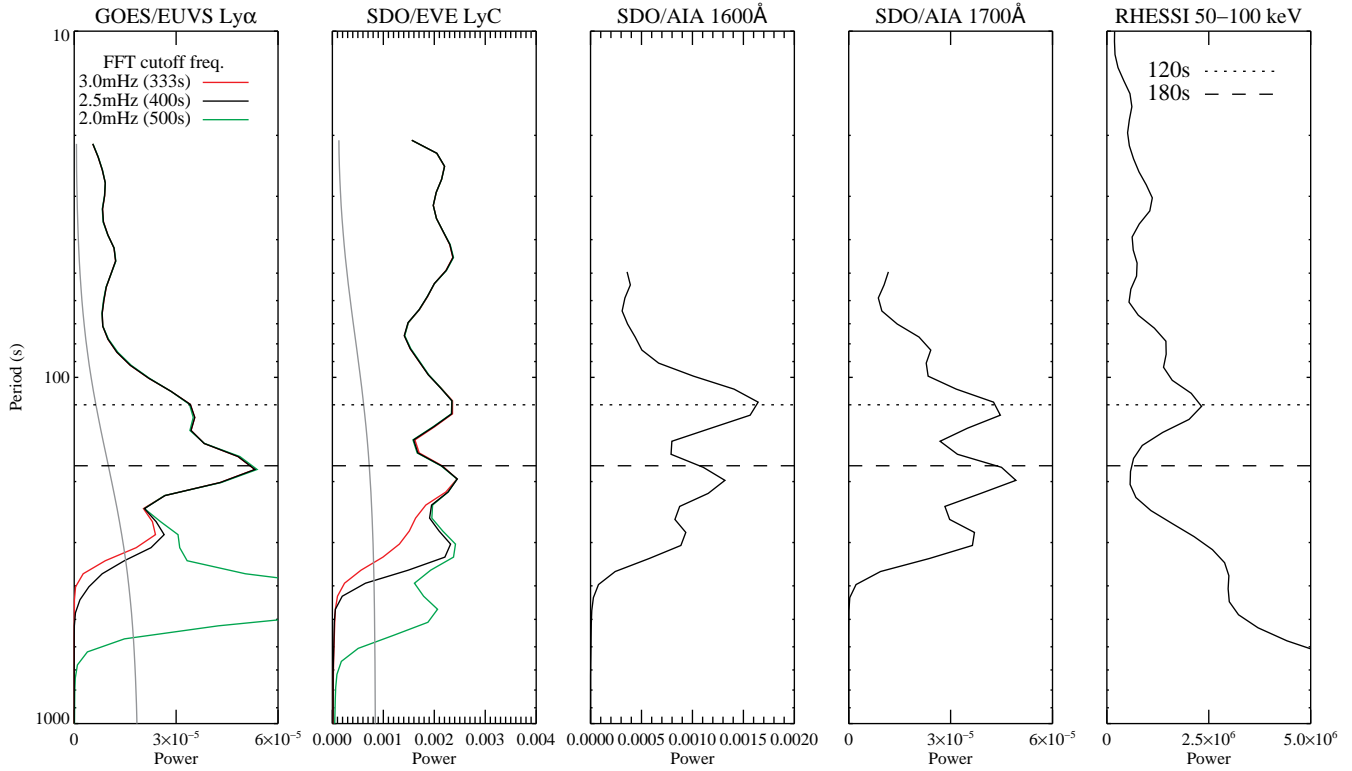}
\caption{From left to right: Global wavelet power spectra for \lya, LyC, 1600\AA, 1700\AA, and 50-100~keV emission, formed by integrating over the duration of the flare (01:30--02:30~UT). The different colored curves in the \lya\ and LyC panels illustrate the resulting spectra obtained using different FFT cutoff frequencies to detrend the data (see legend). The horizontal dotted and dashed lines denote periods of 120s and 180s, respectively. The grey curves in the \lya\ and LyC panels denote the 99\% significance level.}
\label{fig:wave_power}
\end{center}
\end{figure*}

A common explanation for many QPP observations is that they are simply due to bursty energy release and particle acceleration on the measured time scales (e.g. \citealt{bros16}). However, performing a wavelet analysis on the 50--100~keV time profiles from RHESSI for this event did not reveal any power at 3-minute time scales. This not only strengthens the case that the chromospheric pulsations are a genuine oscillatory response that is independent of the energy injection rate, but it also does not support the argument by \cite{sych09} and others that the underlying sunspot oscillation could have been responsible for initiating the energy release and particle acceleration in the first place. Had this been the case then the 3-minute pulsations would have been evident at each step of the energy transport process, including in the HXRs. This is demonstrated by comparing the global wavelet power spectrum (integrated over the duration of the flare; 01:30--02:30~UT) of the HXRs with those of the chromospheric emission, as shown in Figure~\ref{fig:wave_power}. The \lya\ profile shows a strong signature at 180~s (which is independent of the choice of Fourier filter frequency as denoted by the different colored curves), as does LyC to a lesser extent. Both AIA UV channels show enhanced power at around 200~s, while the 50-100~keV emission shows no such power. The two AIA channels, LyC, and \lya\ also all show enhanced power at 120~s to some degree. This {\it does} correspond to the peak in the the HXR power, indicating that the chromospheric response on these timescales is more likely due to accelerated particles. The 3-minute period is a separate phenomenon which is taken to be an oscillatory response of the chromosphere at its acoustic cutoff frequency. \\ \\

\section{Conclusions}
\label{sec:conc}

Observational evidence for 3-minute oscillations in \lya\ and LyC emission from full-disk irradiance measurements during an X-class solar flare is presented. This study supports the notion that when the chromosphere is impulsively disturbed, compressible waves with periods around the acoustic cutoff are generated. It is the recurring compression and expansion of these waves that leads to the oscillation in intensity. The impulsive disturbance may have been caused by the injection of energy, probably in the form of non-thermal electrons, and the amount of the injected energy is likely to have been much larger than that required for sustaining oscillations in non-flaring regions. The oscillation in the flaring region could be identified from the full-disk Lyman alpha and continuum data since the emission from the flaring region is much stronger than the non-flaring regions. \cite{kwak16} reported a similar phenomenon in the chromosphere in response to an impulsive downflow event as observed in chromospheric emission lines by IRIS. 

Such oscillatory responses have been predicted for decades (e.g. \citealt{kalk94}) and this may be the first report of such a disturbance during a major solar flare. Numerical models by \cite{flec91} and \cite{chae15} demonstrate that any impulsive disturbance to a quiescent chromosphere can be found to generate acoustic oscillations around the cutoff frequency; a pre-existing oscillation is not required for a flare-driven wave to exist. However, it is known that such oscillations {\it do} exist around non-flaring active regions as observed in AIA 1600\AA\ and 1700\AA\ images by \cite{rezn12}. Therefore the possibility that the flare somehow ``amplifed'' the quiet-Sun 3-minute oscillations cannot be excluded based on the available data.

It is remarkable that these oscillations show up with a significant amplitude in the full-sun irradiance observations. This may provide a clue to estimate the energy contained in the pulses, assuming that this oscillatory signal comes only from a limited area. Although the 3-minute oscillations appear to be independent of the rate of electron injection, the acoustic waves may transport a significant amount of mechanical energy. \cite{mill14} claimed that the \lya\ line alone can radiate away $\sim$10\% of the non-thermal energy deposited in the chromosphere. \lya\ is also a known driver of disturbances in planetary atmospheres (e.g. the ionospheric D-layer on earth; \citealt{tobi00}), and such oscillations may play a role in changing the atmospheric composition and dynamics during periods of high activity. With \lya\ photometers currently on SDO, GOES13-15, SORCE, the Mars Atmosphere and Volatile EvolutioN (MAVEN; \citealt{epar15}), and scheduled for the next generation of GOES satellites, as well as the \lya\ imager on Solar Orbiter, knowledge of the behavior of this emission during flares could be important when interpreting future science results. \\

\acknowledgments
The authors would like to thank Drs. Mihalis Mathioudakis, Hugh Hudson, and Andrew Inglis, and Laura Hayes, for many insightful discussions on this subject, as well as the anonymous referee for clarification that greatly improved the quality of this manuscript. ROM would like to acknowledge support from NASA LWS/SDO Data Analysis grant NNX14AE07G, and the Science and Technologies Facilities Council for the award of an Ernest Rutherford Fellowship (ST/N004981/1). LF would like to acknowledge support from STFC Consolidated Grants ST/L000741/1 and ST/P000533/1. 


\begin{thebibliography}{36}
\expandafter\ifx\csname natexlab\endcsname\relax\def\natexlab#1{#1}\fi

\bibitem[{{Allred} {et~al.}(2005){Allred}, {Hawley}, {Abbett}, \&
  {Carlsson}}]{allr05}
{Allred}, J.~C., {Hawley}, S.~L., {Abbett}, W.~P., \& {Carlsson}, M. 2005,
  \apj, 630, 573

\bibitem[{{Brosius} {et~al.}(2016){Brosius}, {Daw}, \& {Inglis}}]{bros16}
{Brosius}, J.~W., {Daw}, A.~N., \& {Inglis}, A.~R. 2016, \apj, 830, 101

\bibitem[{{Chae} \& {Goode}(2015)}]{chae15}
{Chae}, J., \& {Goode}, P.~R. 2015, \apj, 808, 118

\bibitem[{{De Pontieu} {et~al.}(2014){De Pontieu}, {Title}, {Lemen}, {Kushner},
  {Akin}, {Allard}, {Berger}, {Boerner}, {Cheung}, {Chou}, {Drake}, {Duncan},
  {Freeland}, {Heyman}, {Hoffman}, {Hurlburt}, {Lindgren}, {Mathur}, {Rehse},
  {Sabolish}, {Seguin}, {Schrijver}, {Tarbell}, {W{\"u}lser}, {Wolfson},
  {Yanari}, {Mudge}, {Nguyen-Phuc}, {Timmons}, {van Bezooijen}, {Weingrod},
  {Brookner}, {Butcher}, {Dougherty}, {Eder}, {Knagenhjelm}, {Larsen},
  {Mansir}, {Phan}, {Boyle}, {Cheimets}, {DeLuca}, {Golub}, {Gates}, {Hertz},
  {McKillop}, {Park}, {Perry}, {Podgorski}, {Reeves}, {Saar}, {Testa}, {Tian},
  {Weber}, {Dunn}, {Eccles}, {Jaeggli}, {Kankelborg}, {Mashburn}, {Pust},
  {Springer}, {Carvalho}, {Kleint}, {Marmie}, {Mazmanian}, {Pereira}, {Sawyer},
  {Strong}, {Worden}, {Carlsson}, {Hansteen}, {Leenaarts}, {Wiesmann},
  {Aloise}, {Chu}, {Bush}, {Scherrer}, {Brekke}, {Martinez-Sykora}, {Lites},
  {McIntosh}, {Uitenbroek}, {Okamoto}, {Gummin}, {Auker}, {Jerram}, {Pool}, \&
  {Waltham}}]{depo14}
{De Pontieu}, B., {et~al.} 2014, \solphys, 289, 2733

\bibitem[{{Dennis} {et~al.}(2017){Dennis}, {Tolbert}, {Inglis}, {Ireland},
  {Wang}, {Holman}, {Hayes}, \& {Gallagher}}]{denn17}
{Dennis}, B.~R., {Tolbert}, A.~K., {Inglis}, A., {Ireland}, J., {Wang}, T.,
  {Holman}, G.~D., {Hayes}, L.~A., \& {Gallagher}, P.~T. 2017, \apj, 836, 84

\bibitem[{{Dolla} {et~al.}(2012){Dolla}, {Marqu{\'e}}, {Seaton}, {Van
  Doorsselaere}, {Dominique}, {Berghmans}, {Cabanas}, {De Groof}, {Schmutz},
  {Verdini}, {West}, {Zender}, \& {Zhukov}}]{doll12}
{Dolla}, L., {et~al.} 2012, \apjl, 749, L16

\bibitem[{{Eparvier} {et~al.}(2015){Eparvier}, {Chamberlin}, {Woods}, \&
  {Thiemann}}]{epar15}
{Eparvier}, F.~G., {Chamberlin}, P.~C., {Woods}, T.~N., \& {Thiemann}, E.~M.~B.
  2015, \ssr, 195, 293

\bibitem[{Fischler \& Bolles(1981)}]{fisc81}
Fischler, M.~A., \& Bolles, R.~C. 1981, Communications of the ACM, 24, 381

\bibitem[{{Fleck} \& {Schmitz}(1991)}]{flec91}
{Fleck}, B., \& {Schmitz}, F. 1991, \aap, 250, 235

\bibitem[{{Hanser} \& {Sellers}(1996)}]{hans96}
{Hanser}, F.~A., \& {Sellers}, F.~B. 1996, in \procspie, Vol. 2812, GOES-8 and
  Beyond, ed. E.~R. {Washwell}, 344--352

\bibitem[{{Hayes} {et~al.}(2016){Hayes}, {Gallagher}, {Dennis}, {Ireland},
  {Inglis}, \& {Ryan}}]{haye16}
{Hayes}, L.~A., {Gallagher}, P.~T., {Dennis}, B.~R., {Ireland}, J., {Inglis},
  A.~R., \& {Ryan}, D.~F. 2016, \apjl, 827, L30

\bibitem[{{Inglis} {et~al.}(2016){Inglis}, {Ireland}, {Dennis}, {Hayes}, \&
  {Gallagher}}]{ingl16}
{Inglis}, A.~R., {Ireland}, J., {Dennis}, B.~R., {Hayes}, L., \& {Gallagher},
  P. 2016, \apj, 833, 284

\bibitem[{{Kalkofen} {et~al.}(1994){Kalkofen}, {Rossi}, {Bodo}, \&
  {Massaglia}}]{kalk94}
{Kalkofen}, W., {Rossi}, P., {Bodo}, G., \& {Massaglia}, S. 1994, \aap, 284,
  976

\bibitem[{{Kumar} {et~al.}(2016){Kumar}, {Nakariakov}, \& {Cho}}]{kuma16}
{Kumar}, P., {Nakariakov}, V.~M., \& {Cho}, K.-S. 2016, \apj, 822, 7

\bibitem[{{Kwak} {et~al.}(2016){Kwak}, {Chae}, {Song}, {Kim}, {Lim}, \&
  {Madjarska}}]{kwak16}
{Kwak}, H., {Chae}, J., {Song}, D., {Kim}, Y.-H., {Lim}, E.-K., \& {Madjarska},
  M.~S. 2016, \apjl, 821, L30

\bibitem[{Lamb(1909)}]{lamb09}
Lamb, H. 1909, Proceedings of the London Mathematical Society, s2-7, 122

\bibitem[{{Lemen} {et~al.}(2012){Lemen}, {Title}, {Akin}, {Boerner}, {Chou},
  {Drake}, {Duncan}, {Edwards}, {Friedlaender}, {Heyman}, {Hurlburt}, {Katz},
  {Kushner}, {Levay}, {Lindgren}, {Mathur}, {McFeaters}, {Mitchell}, {Rehse},
  {Schrijver}, {Springer}, {Stern}, {Tarbell}, {Wuelser}, {Wolfson}, {Yanari},
  {Bookbinder}, {Cheimets}, {Caldwell}, {Deluca}, {Gates}, {Golub}, {Park},
  {Podgorski}, {Bush}, {Scherrer}, {Gummin}, {Smith}, {Auker}, {Jerram},
  {Pool}, {Soufli}, {Windt}, {Beardsley}, {Clapp}, {Lang}, \&
  {Waltham}}]{leme12}
{Lemen}, J.~R., {et~al.} 2012, \solphys, 275, 17

\bibitem[{{Li} {et~al.}(2015){Li}, {Ning}, \& {Zhang}}]{li15}
{Li}, D., {Ning}, Z.~J., \& {Zhang}, Q.~M. 2015, \apj, 807, 72

\bibitem[{{Lin} {et~al.}(2002){Lin}, {Dennis}, {Hurford}, {Smith}, {Zehnder},
  {Harvey}, {Curtis}, {Pankow}, {Turin}, {Bester}, {Csillaghy}, {Lewis},
  {Madden}, {van Beek}, {Appleby}, {Raudorf}, {McTiernan}, {Ramaty}, {Schmahl},
  {Schwartz}, {Krucker}, {Abiad}, {Quinn}, {Berg}, {Hashii}, {Sterling},
  {Jackson}, {Pratt}, {Campbell}, {Malone}, {Landis}, {Barrington-Leigh},
  {Slassi-Sennou}, {Cork}, {Clark}, {Amato}, {Orwig}, {Boyle}, {Banks},
  {Shirey}, {Tolbert}, {Zarro}, {Snow}, {Thomsen}, {Henneck}, {McHedlishvili},
  {Ming}, {Fivian}, {Jordan}, {Wanner}, {Crubb}, {Preble}, {Matranga}, {Benz},
  {Hudson}, {Canfield}, {Holman}, {Crannell}, {Kosugi}, {Emslie}, {Vilmer},
  {Brown}, {Johns-Krull}, {Aschwanden}, {Metcalf}, \& {Conway}}]{lin02}
{Lin}, R.~P., {et~al.} 2002, \solphys, 210, 3

\bibitem[{{Milligan} \& {Chamberlin}(2016)}]{mill16}
{Milligan}, R.~O., \& {Chamberlin}, P.~C. 2016, \aap, 587, A123

\bibitem[{{Milligan} {et~al.}(2012){Milligan}, {Chamberlin}, {Hudson}, {Woods},
  {Mathioudakis}, {Fletcher}, {Kowalski}, \& {Keenan}}]{mill12}
{Milligan}, R.~O., {Chamberlin}, P.~C., {Hudson}, H.~S., {Woods}, T.~N.,
  {Mathioudakis}, M., {Fletcher}, L., {Kowalski}, A.~F., \& {Keenan}, F.~P.
  2012, \apjl, 748, L14

\bibitem[{{Milligan} \& {Dennis}(2009)}]{mill09}
{Milligan}, R.~O., \& {Dennis}, B.~R. 2009, \apj, 699, 968

\bibitem[{{Milligan} {et~al.}(2014){Milligan}, {Kerr}, {Dennis}, {Hudson},
  {Fletcher}, {Allred}, {Chamberlin}, {Ireland}, {Mathioudakis}, \&
  {Keenan}}]{mill14}
{Milligan}, R.~O., {et~al.} 2014, \apj, 793, 70

\bibitem[{{Nakariakov} \& {Melnikov}(2009)}]{naka09}
{Nakariakov}, V.~M., \& {Melnikov}, V.~F. 2009, \ssr, 149, 119

\bibitem[{{Ning}(2017)}]{ning17}
{Ning}, Z. 2017, \solphys, 292, 11

\bibitem[{{Noyes} \& {Leighton}(1963)}]{noye63}
{Noyes}, R.~W., \& {Leighton}, R.~B. 1963, \apj, 138, 631

\bibitem[{{Pesnell} {et~al.}(2012){Pesnell}, {Thompson}, \&
  {Chamberlin}}]{pesn12}
{Pesnell}, W.~D., {Thompson}, B.~J., \& {Chamberlin}, P.~C. 2012, \solphys,
  275, 3

\bibitem[Reznikova et al.(2012)]{rezn12} Reznikova, V.~E., Shibasaki, K., Sych, R.~A., \& Nakariakov, V.~M.\ 2012, \apj, 746, 119 

\bibitem[{{Sim{\~o}es} {et~al.}(2015){Sim{\~o}es}, {Hudson}, \&
  {Fletcher}}]{simo15}
{Sim{\~o}es}, P.~J.~A., {Hudson}, H.~S., \& {Fletcher}, L. 2015, \solphys, 290,
  3625

\bibitem[{{Sutmann} {et~al.}(1998){Sutmann}, {Musielak}, \&
  {Ulmschneider}}]{sutm98}
{Sutmann}, G., {Musielak}, Z.~E., \& {Ulmschneider}, P. 1998, \aap, 340, 556

\bibitem[{{Sych} {et~al.}(2009){Sych}, {Nakariakov}, {Karlicky}, \&
  {Anfinogentov}}]{sych09}
{Sych}, R., {Nakariakov}, V.~M., {Karlicky}, M., \& {Anfinogentov}, S. 2009,
  \aap, 505, 791

\bibitem[{{Tobiska} {et~al.}(2000){Tobiska}, {Woods}, {Eparvier}, {Viereck},
  {Floyd}, {Bouwer}, {Rottman}, \& {White}}]{tobi00}
{Tobiska}, W.~K., {Woods}, T., {Eparvier}, F., {Viereck}, R., {Floyd}, L.,
  {Bouwer}, D., {Rottman}, G., \& {White}, O.~R. 2000, Journal of Atmospheric
  and Solar-Terrestrial Physics, 62, 1233

\bibitem[{{Torrence} \& {Compo}(1998)}]{torr98}
{Torrence}, C., \& {Compo}, G.~P. 1998, Bulletin of the American Meteorological
  Society, 79, 61

\bibitem[{{Van Doorsselaere} {et~al.}(2016){Van Doorsselaere}, {Kupriyanova},
  \& {Yuan}}]{vand16}
{Van Doorsselaere}, T., {Kupriyanova}, E.~G., \& {Yuan}, D. 2016, \solphys

\bibitem[{Viereck {et~al.}(2007)Viereck, Hanser, Wise, Guha, Jones, McMullin,
  Plunket, Strickland, \& Evans}]{vier07}
Viereck, R., {et~al.} 2007, Solar Physics and Space Weather Instrumentation II.
  Edited by Fineschi, 6689, 66890K

\bibitem[{{Woods} {et~al.}(2012){Woods}, {Eparvier}, {Hock}, {Jones},
  {Woodraska}, {Judge}, {Didkovsky}, {Lean}, {Mariska}, {Warren}, {McMullin},
  {Chamberlin}, {Berthiaume}, {Bailey}, {Fuller-Rowell}, {Sojka}, {Tobiska}, \&
  {Viereck}}]{wood12}
{Woods}, T.~N., {et~al.} 2012, \solphys, 275, 115

\end{thebibliography}
\end{document}